\documentclass[12pt]{article}
\usepackage{amssymb, url, amsthm, amsmath}

\usepackage{textcomp}
\def\boxit#1{%
{\hbox{\lower3pt\hbox{\vrule\vbox{\hrule\kern2pt%
\hbox{\kern2pt$#1$\kern2pt}\kern2pt\hrule}\vrule}}}}
\bigskip
\bigskip
\def\be{\begin{equation}}
\def\ee{\end{equation}}

\def\R{{\sf I\kern-.2em R}}
\def\N{{\sf I\kern-.2em N}}
\def\C{\kern.1em{\raise.47ex\hbox{$\scriptscriptstyle
$}}\kern-.40em{\sf C}}
\def\Z{{\sf Z\kern-.32em Z}}

\def\hat{\widehat}

\def\hat{\widehat}

\newtheorem{theorem}{\noindent Theorem}

\newtheorem{definition}{\noindent Definition}
\newtheorem{corollary}{\noindent Corollary}

\newtheorem{statement}{\noindent Proposition}
\newtheorem{remark}{\noindent Remark}

\urldef\vershik\url{vershik@pdmi.ras.ru}

\author {A.~M.~Vershik\thanks{%
Steklov Mathematical Institute at St.~Petersburg.
E-mail: \vershik.
Partly supported by the grants RFBR 05-01-0089 and CRDF RUM1-2622-ST-04.%
}}

\date{28.08.07}

\title{Does there exist the Lebesgue measure in the infinite-dimensional space?}

\begin{document}
 \maketitle
\rightline{\it To V.~I.~Arnold with profound respect.}

\begin{abstract}
 We consider the sigma-finite measures
 in the space of vector-valued distributions on the manifold  $X$ with characteristic functional
 $$\Psi(f)=\exp\{-\theta\int_X\ln||f(x)||dx\},\ \theta>0.$$
 The collection of such measures constitutes a one-parameter semigroup relative to $\theta$.
 In the case of scalar distributions and $\theta=1$, this measure may be called
 the {\it infinite-dimensional Lebesgue measure.}
 We consider the weak limit of Haar measures on the Cartan subgroup of
 the group $SL(n,{\Bbb R})$ when $n$ tends to infinity.
  The measure in the limit is invariant under the linear action of some infinite-dimensional
  Abelian group which is an analog of an infinite-dimensional Cartan
  subgroup. This fact can be a justification of the name
 \emph{Lebesgue} as a valid name for the measure in question.
Application to the representation theory of the current groups was
one of the reason to define this measure. The measure also is
closely related to the Poisson--Dirichlet measures well known in
combinatorics and probability theory.

 The only known example of the analogous asymptotical behavior of the uniform measure on the
 homogeneous manifold is classical Maxwell-Poincar\'e lemma which asserts that
 the weak limit of uniform measures on the Euclidean sphere of appropriate radius as dimension
 tends to infinity is the standard infinite-dimensional Gaussian measure or white noise.
 Our situation is similar but all the measures are no more finite
 but sigma-finite.
 \end{abstract}

\tableofcontents

\section{Introduction}
\subsection{On asymptotic approach to measure and integration in infinite-dimensional spaces}

In his remarkable but less known, compared with other works, paper
``Approximative properties of matrices of high finite order''
(\cite{vN}), J.~von Neumann wrote that experts in functional
analysis neglect problems concerning spaces of high finite dimension
in favor of the study of actually infinite-dimensional spaces.
Possibly, in the last third of the 20th century the situation has
slightly changed, but one still cannot say that we understand
analysis in the spaces of dimension, say, $10^{24}$ better than in
the infinite-dimensional Hilbert space (where ``almost everything is
clear''!) \footnote{A subheading of Chapter~5 in the E.~Borel's book
\cite{Bor1} reads: ``Functions in high number of variables: areas
and volumes in the geometry of $10^{24}$-dimensional spaces''.}
Specifically, this is true in what concerns problems in measure
theory and integration in infinite-dimensional spaces. Never-ceasing
attempts to justify the notion of Feynman integral, which is so
important to physicists, and to embed it into one or another general
scheme of integration over a measure do not evoke interest or
approval of physicists and apprehension of mathematicians. It is
easy to understand this lack of enthusiasm: physical modelling is
always or almost always based upon asymptotic constructions (in
dimension, number of particles, some constants, etc.). On the
contrary, mathematicians usually try to interpret these
constructions as actually infinite (infinite-dimensional). This is
productive and necessary within some limits but inevitably results
in certain difficulty of interpretation when one tries to absolutize
the limiting constructions. Certainly, it is impossible to say that
asymptotic approach can be a substitute of actually infinite
constructions, and there is no need in such substitution. It is
important to understand what effects, in the infinite-dimensional
case, really survive, or grow out of asymptotic finite-dimensional
properties, and how to obtain them. We will investigate an example
of asymptotic behavior of measures on classical homogeneous spaces
which leads to a remarkable limiting measure (to be precise, a
one-parameter family of measures), the Lebesgue measure in the
infinite-dimensional space. This measure (in different, actually
infinite terms) was earlier discovered in connection with
representation theory of current groups \cite{GGV}. The role it
plays in combinatorics and representation theory is probably not
smaller than that of the Gaussian measure. Its properties and deep
connections with, for instance, Poisson--Dirichlet measures are
probably covered in this work for the first time (cf. \cite{TVY})
and need further investigation.

\subsection{About this paper}
We begin Subsection~2 with the classical and well-known calculation,
the so-called ``Poincar\'e's Lemma'' that substantiates the Maxwell
(Gaussian) distribution of velocities in statistical physics. This
example shows how the infinite-dimensional Gaussian measure (``white
noise'') arises as a limiting distribution of the radius-vector of a
point on the Euclidean sphere as the dimension and the radius of the
sphere tend to infinity coherently. The aim of the present paper is
to demonstrate that, as in the above-mentioned example, which will
systematically play the role of a reference example for the main
theme of the paper, there exists another series of homogeneous
spaces of the Cartan subgroup in $SL(n,\Bbb R)$, on which the
invariant measures, as in the case of Maxwell-Poincar\'e's Lemma,
weakly converge to quite another, now sigma-finite, measure which
reminds the infinite-dimensional Lebesgue measure. The symmetry
group of the measure in question is as large as in the case of the
Gaussian measure, but quite different one. This measure is related
to the remarkable Poisson--Dirichlet measures of combinatorial
origin. There is reason to compare the Wiener and our Lebesgue
measures: they can be viewed as the measures corresponding to the
extreme values in the segment $\alpha \in [0,2]$ whose inner points
parameterize L\'evy measures of stable laws; to be more precise, our
measure is the derivative of these measures over $\alpha$ at the
point 0. Apart from interest per se, the measures in question are
used (and first appeared) in representation theory of current
groups. However, our main aim is the description of these measures
in the geometric and asymptotic aspects. In Sect.~2, we give proofs
of the Maxwell-Poincar\'e Lemma which we use for comparison in many
situations. This comparison is useful and allows us to outline
further natural generalizations. We comment this lemma from various
points of view.

In Sect.~3, we consider the orbits of the Cartan subgroup $SDiag(n,\Bbb R)$ in the group
$SL(n,\Bbb R)$. It is convenient to start with the study of the positive part
$SDiag_+(n \Bbb R)$ of the Cartan subgroup and of its orbit, postponing the general case to
Sect.~4. Further, we embed the orbits into the cone $K^+$ of positive step functions
on the segment and define the weak convergence of the invariant measures on these orbits as
the convergence of their Laplace transforms. The limit of the Laplace transformations of the
properly normalized measures on $SDiag_+(n \Bbb R)$ is the functional
 $$\Phi_{\theta}(f)=\exp\Big\{-\theta \int_X \ln f(x)dx\Big\}, \quad \theta>0,$$
and we use several methods of finding it. This functional is defined on the set of functions
whose logarithm has finite integral; it is invariant under all changes of variables keeping the
measure invariant and under multiplication by functions whose logarithm has zero integral.

The main object of Sect.~4 is an explicit definition of the
sigma-additive sigma-finite measure ${\cal L}_{\theta}^+$. First we
define the weak distribution $\Xi_{\theta}$ (Subsect.~4.1) on the
cone $K$ whose Laplace transform is $\Phi_{\theta}$. Then we
introduce the cone $D_+$ of discrete positive measures of finite
mass defined on $X$, which is in duality with the cone $K^+$. Thus
the weak distribution $\Xi_{\theta}$ may be viewed as a pre-measure
on $D_+$. We emphasize once more that our object is not finite but
infinite weak distributions and measures. So the usual tools like
projections, etc., cannot be applied here. The final step of the
construction is the proof of the existence of a true sigma-additive
measure that is a continuation of our weak distribution. This is
done in a constructive way using an infinite (``poissonized'', or
conic) version of the Poisson--Dirichlet measures $PD(\theta),\;
\theta>0,$ which became popular in the last years. These measures
are defined on the simplex of monotone positive series with sum one;
we describe them in Appendix~1. We need their sigma-finite versions,
$PDC$, the ``conic Poisson--Dirichlet measures'', which are defined
on the cone of monotone convergent positive series. These measures
are direct products of the Poisson--Dirichlet measure $D(\theta)$
and the measure on the half-line $L_{\theta}$ defined by the density
$t^{\theta-1}/\Gamma(\theta)$. (The measure on the half-line is a
``distribution'' of the sum of the series.)

In Subsection~4.2, which plays a central role in our exposition, we define the principal object,
the multiplicative measures ${\cal L}_{\theta}$, as
\emph{an image of the product of the Bernoulli measures
$m^{\infty}$ and the conic Poisson--Dirichlet measures $PDK(\theta)$} described above.

These measures are eventually the weak limits of the measures defined on the sequence of
$SDiag_+(n, \Bbb R)$-orbits. The measure corresponding to $\theta=1$ is called the multiplicative
Lebesgue measure on the cone  $D_+$.

Thus, our scheme of the introducing the multiplicative measures, the Lebesgue measure in
particular, is the following.

\emph{We define the measures on the orbits of the Cartan subgroup, then find the limit of their
Laplace transform; the latter is the Laplace transform of some weak distribution and we define
the measure, which is a continuation of this distribution, by taking an explicit image of
the conic Poisson--Dirichlet measures multiplied by the Bernoulli product measure}.

In this apparently long way the concluding step does not depend on the preceding ones. This allows
one to introduce the measures we are looking for directly, independently of the preceding steps.
However, this economy of efforts conceals the asymptotic and geometric sense of the measure constructed.
The reader who does not care of this sense can pass to the Subsection~4.2 immediately after
the introduction.

We summarize the properties of these measures.
\begin{enumerate}
    \item They are the weak limits of the measures on orbits of the positive
    Cartan subgroups;
    \item Their Laplace transform is
    $$\Phi_{\theta}(f)=\exp\Big\{-\theta \int \ln f(x)dx\Big\},\quad \theta>0;$$

    \item They are the images of the product of the Poisson--Dirichlet measures on the
    simplex of positive convergent series summing to one by the Bernoulli measure and
    the Lebesgue measure on the half-line.
\end{enumerate}

On the other hand, these measures behave like the laws of L\'evy processes, but with infinite probability:
our measures are absolutely continuous and even equivalent to the laws of L\'evy gamma processes
on subordinators. Exactly in this way they were defined in \cite{TVY}, and eventually in this
way they were discovered in \cite{VGG1,VGG2}. In Subsection~4.3, we connect these measures
to L\'evy gamma processes, subordinated or complete. In \cite{GGV,TVY}, an opposite way
to define the measures is adopted. They are defined via a gamma process by the introduction
of densities. This method is less analytic and transparent, especially in the infinite-dimensional
case.

In Subsection~4.4, we give an additive version of the description of these measures and
show that they present the first example of a sigma-finite measure invariant under shifts
by vectors of an infinite-dimensional Banach space.

Further, in Subsection~5.1, the main definition and all the other definitions are repeated
in the case of signed measures; the cone is replaced by the vector space $D$, positive
series by absolutely convergent ones, etc. This transition is easy, and the most important properties
are already visible in the ``positive'' version. These extended and most important
measures have the following properties.

\begin{enumerate}
    \item They are the weak limits of measures on the orbits of the complete Cartan subgroup.
    \item Their Laplace transform is
    $$\Phi_{\theta}(f)=\exp\Big\{-\theta \int \ln |f(x)|dx\Big\},\quad \theta>0$$
    (the logarithm is replaced by the logarithm of the modulus).
    \item They are the images of the products of extended Poisson--Dirichlet measures on
    the octahedron composed by all decreasing (in modulus) absolutely convergent
    series, a Bernoulli measure, and the Lebesgue measure on the line.
    \item Finally (the most important): \emph{these measures are invariant relative to the group
    of multiplicators by the functions $f$ with zero integral of\; $\log|f|$,
    they are projectively invariant relative the multiplication by the functions $f$ with
    finite integral of\; $\log|f|$, and (Subsection~5.2) they are invariant under the changes of variables that
    leave the measure invariant}.\footnote{I.e., they are invariant relative to the normalizer
    of the infinite-dimensional torus (= the group of multiplication operators).}
\end{enumerate}

In Subsection~5.3, we remind the connection of these measures with representation theory
of current groups. Finally, in Subsection~5.4 we define a generalization of the Lebesgue
and the Poisson--Dirichlet measures to the vector case, which is necessary for the representations
of current groups with coefficients in the group $SO(n,1)$.

The first appendix contains the most important information about the Poisson--Dirichlet measures
and their applications in probability, algebra, and number theory. In the second appendix we
discuss the conditions that are imposed on the group of admissible shifts by the properties of
invariance and quasi-invariance of the measures under this group, and we explain what
is new in the additive approach to infinite-dimensional Lebesgue measures introduced
here.

\section{A brief historic digression: white noise according to Maxwell--Poincar\'e--%
Borel, and commentaries}

\subsection{Maxwell-Poincar\'e's Lemma}

A remarkable example of asymptotic approach to infinite-dimensional
objects is presented by the following way to introduce the
Maxwell-Boltzmann distribution in mathematical physics. Consider the
small canonical ensemble of the velocities of a system of identical
particles with energy
$$H(v_1,\dots v_n)=\frac{1}{2} \sum _k||v_k||^2.$$
Since we do not care about the dimension, the velocities may be treated as scalars ($d=1$).
A natural measure carried by the small ensemble is the normalized Lebesgue measure on the corresponding
Euclidean sphere (because the measure must be orthogonally invariant). On the other hand, consider
the canonical ensemble of velocities with Gibbs measure, i.e., the measure with density
$\exp\{-сH(v_1, \dots v_n)\},\; c>0$, on it. When normalized, it becomes the standard
Gaussian measure. Then we increase the number of particles and, simultaneously, the total
energy. The question is: do the asymptotic distributions in both ensembles coincide? The answer
is contained in the following beautifully simple fact that can be formulated, in the current
terms, as follows.

\begin{theorem}
Consider the sequence of the normalized Lebesgue measures on the Euclidean spheres
$S_{r_n}^{n-1}\subset {\Bbb R}^n$ of radius $r_n=c\sqrt n,\; c>0$, and the limit
of spaces
 $${\Bbb R}^1\subset {\Bbb R}^2 \subset \dots \subset {\Bbb R}^n
 \subset \dots \subset {\Bbb R}^{\infty}.$$
Then the weak limit of these measures is the standard Gaussian measure $\mu$
which is the infinite product of the identical Gaussian measures on the line with zero mean and
variance $c^2$. It is clear that the sequence of Gibbs measures has the same weak limit.
\end{theorem}
Thus the infinite-dimensional ensemble that is the limit of both canonical ensembles
in the above-described sense exists.
\begin{proof}
The weak convergence of a sequence of measures is, by definition, the convergence of
the corresponding sequence of finite-dimensional distributions for any finite collection
of linear functionals. In its turn, it is sufficient for this that the distribution of
a single (arbitrary) functional converge; for instance, one can consider the functional that
takes the first coordinate
of a vector. As a result, the question reduces to the following calculation. One should
find the limiting distribution of the projection of the Lebesgue measure on the sphere
$S_{r_n}^{n-1}$ onto the first coordinate. The density relative to the Lebesgue measure
of the projection of the (normalized) such measure is $C_n(r_n^2-x^2)^{\frac{n-2}{2}}$.
After an evident renormalization, as $ r/\!\sqrt n \to \theta >0$, we get the density
$C\exp(-\theta x^2),\; \theta>0,$ of the Gaussian measure as a limit.
\end{proof}

The same result can be obtained in a number of different ways. For example, one can use
the Fourier transform and consider the asymptotic behavior of Bessel functions. For the
sphere $S^{n-1}$, set  $\nu=(n-2)/2$. From \cite{RG}, formula 3.771.8
$$\int_0^r(r^2-x^2)^{\nu -1/2}\exp(itx)dx=C\cdot\left(\frac{2r}{t}\right)^{\nu}J_{\nu}(tr),$$
where $C=\frac{\sqrt \pi/2}{\Gamma(\nu +1/2)}$,
and the formula giving the asymptotical behavior of the Bessel function $J_{\nu}(.)$
as its argument and number $\nu$ tend to infinity, it follows that
$$\lim_{n\to \infty}
\int_{S^{n-1}_{r_n}}\exp(it\omega)d\Omega_n(\omega)=\exp(-\theta t^2)$$
as $r/\sqrt n =r/\sqrt{2(\nu+1)} \to \theta>0$, where $\Omega_n$ is the normalized
Lebesgue measure on the sphere $S^{n-1}_{r_n}$. Thus the sequence of the Fourier transforms
tends to the Fourier transform of the Gaussian measure, and the weak convergence of the measure
follows. We give an analog of this very proof in the situation in question replacing
Fourier transform with Laplace transform.

\subsection{Comments}

\paragraph{1.} A more serious comprehension of the latter calculation is the following. This demonstration
can be viewed as the derivation of the empirical distribution of the first (and then any) coordinate
of the vector in the space ${\Bbb R}^{\infty}$ relative to an a priory unknown spherically invariant
measure. Indeed, it follows from the general ergodic and martingale convergence theorems (see the
so-called ergodic method in \cite{V1,V3}) that the limit of such functional is the limit of its
empirical distributions
for any probability Borel ergodic measure in the space ${\Bbb R}^{\infty}$ that is invariant
under the action of all finite-dimensional orthogonal (in the $l^2$ sense) groups (and,
consequently, under the whole infinite-dimensional orthogonal group $O(\infty)$ in $l^2$). But the
thing is that we do not know in the beginning what set of vectors constitutes the set of ``almost all''
vectors relative to the measure we are looking for, and so we do not know what orbits to take.
However, the theorems cited imply that taking all the orbits we will not miss any invariant
ergodic measure. It turns out that in our case it suffices to take orbits having the form
$$(\underbrace{x,x,\dots x}_n, 0,0 \dots),$$ these and only these orbits give all necessary measures,
the other sequences of orbits do not have nontrivial limits. This is the manifestation of the
fact that the average square of the norm of such vector relative to the Gaussian measure
grows proportionally to $n$, and consequently there are no ergodic measures except the
Gaussian ones. It is clear that the knowledge of the distributions of all (in our case, one)
linear functionals defines the measure completely.

It immediately follows that the general spherically invariant measure is a mixture of
Gaussian measures with various dispersions, i.e., the general form of the
characteristic functional of a spherically invariant measure is the following:
$\int_0^{\infty}\exp(-cx^2)dm(c)$. Hence the Schoenberg theorem follows which states that all
indecomposable positive definite normalized functions of the norm of a vector in an
infinite-dimensional
Hilbert space have the form $\phi(h)=\exp(-||h||^2)$. This fact, which is essentially one of the
versions of the ergodic theorem (or the martingale theorem), makes it possible to describe all
invariant measures, not only in this particular example but also in the general case, by choosing
in a special way the orbits of the subgroups that approximate the given group. This is
essentially what we do in the example of noncompact Cartan subgroups, where we also describe
all invariant measures.

\paragraph{2.} A more delicate fact, which we will use below, is that the action of the whole
infinite-dimensional orthogonal group $O^{\infty}$ in the space ${\Bbb R}^{\infty}$ should be
meant only in the sense that every orthogonal operator $g \in O^{\infty}$ is defined,
and acts leaving the Gaussian measure invariant, on a \emph{certain} measurable linear
subspace of total measure (it can be easily constructed using, for example, the spectral
decomposition of  $g$ in $l^2$) that \emph{depends on the operator}, but a
common \emph{linear measurable subspace} where all orthogonal operators were defined
simultaneously does not exist, as was proved in $\cite{V2}$. It was also shown recently
in $\cite{GTW}$ that no measurable set of total measure exist where all the elements
of the group $O^{\infty}$ were defined simultaneously.
\footnote{It was not mentioned in $\cite{GTW}$ that the absence of common \emph{linear subspace}
was proved in $\cite{V2}$.}
This gives an example of the group action that does not admit an individual measurable
realization. It is well known that in the case of locally compact groups a measurable
realization always exists.

\paragraph{3.} One can define a measure invariant under arbitrary group possessing a dense subgroup that is
a union of an increasing sequence of compact or locally compact
subgroups in a similar manner (this is the ergodic method of the
description of invariant measures, characters, etc.) We choose an
orbit for any subgroup from the given sequence of groups and take an
invariant measure on the orbit. Then we look for all cases when
these measures on the orbits weakly converge. The ergodic theorem or
the martingale convergence theorem guarantee that the list of
invariant measures thus obtained is complete. The case of compact
groups is simpler. For the Maxwell-Poincar\'e-- case, the orbits are
$n$-dimensional spheres of radius $c\sqrt n$, and the Lebesgue
measures on them weakly converge to the Gaussian measure. Exactly in
the same way, changing the spheres and embedding maps, one can
obtain any Gaussian measure in the infinite-dimensional space, since
they all are linearly isomorphic. For example, white noise as
generalized in the sense of Gelfand-Ito gaussian process or more
exactly, the corresponding gaussian measure in the space of Schwartz
distributions can be constructed in this manner as a weak limit of
the sequence of uniform measures on the unit (in the $L^2$-norm)
spheres on finite dimensional subspaces. We will use the described
technique for noncompact groups in what follows.

\paragraph{4.}
Some remarks of historical character. The above calculation can be
found in many books and papers. Most commonly, it is called
Poincar\'e's Lemma, or even the Maxwell Theorem \cite{Max}, (make
sense to mention also the name L.Boltzmann - Maxwell-Boltzmann
distribution). Yet a number of authors \cite{Str,Diac} claim that
they could not find this lemma nowhere in the papers by Poincar\'e.
E.~Borel quotes it many times \cite{Bor1, Bor2}; however, he does
not mention Poincar\'e in this connection while abundantly quoting
him on many other occasions \cite{Poin}. D.~Strook, G.~McKean and
M.~Yor \cite{Yor} showed me a paper \cite{Meh} (1866) by the German
mathematician F.~Mehler where on can already find this calculation;
It seems that E.~Borel did not know about this work. In fact, there
is a theorem in \cite{Meh} that the generating function of spherical
harmonics converges, as its index increases, to the generating
function of Hermite polynomials. This evidently implies our modest
fact (and even the convergence of all the moments of the
distributions); however, the geometrical picture that is the essence
of the method remains concealed in this general theorem. H.~McKean
informed me that, among the others, M.~Kac mentioned H.~Poincar\'e
as the author of this statement. See also the recent preprint by
P.~Cartier \cite{Car}. One can guess that H.~Poincar\'e mentioned
this method of obtaining Maxwell's distribution in his lectures but
has not written it down: the fact that he was aware of this
calculation can be seen from his lectures \cite{Poin}. Thus,
according to the principle expressed by many authors (some of whom,
following this very principle, attribute the principle itself to
V.~I.~Arnold) which states that the names ascribed by the later
generations to theories, theorems, lemmas rarely belong to the true
discoverers of these theories etc., we continue to call the
statement in question Maxwell-Poincar\'e's Lemma, taking a risk to
violate the (possibly erroneous) tradition.

In the present paper we show that in another, non-compact,
sigma-finite version, the analogous asymptotic method brings us not
to the Gaussian measure, but to a no less remarkable
infinite-dimensional measure. It appeared earlier in representation
theory of the current group \cite{VGG2} and, as it turned out later,
is closely related to the L\'evy gamma process. We will describe it
in various aspects but will show what is the most natural way to
discover it using geometric approach.

\section{A measures on the orbits of the Cartan subgroups and the weak
limits of its Laplace transform}

\subsection{The orbits of the Cartan subgroups.}

Instead of $(n-1)$-dimensional spheres $S^{n-1}_r$ of radius
$r_n=c\sqrt n$ in Maxwell-Poincar\'e's Lemma, we consider the
hypersurfaces in ${\Bbb R}^n$ \footnote{sometimes this hyperspheres
called affine spheres} :
$$
M^{n-1}_{r_n} = \bigg\{(y_1, \dots y_n): \prod_{k=1}^n y_k =r_n^n
>0;\; y_k>0 k=1 \dots n \bigg\}
$$
 The number $r_n$ will be called the \emph{radius} of the
 hypersurface,
- it depends on $n$, - and will be specified later. On this
hypersurface $M^{n-1}_r$ (for all $r$), the group $SDiag_+(n,\Bbb
R)$ of positive diagonal matrices with determinant one, i.e., the
positive part of the Cartan subgroup of the group $SL(n,\Bbb R)$,
acts freely and transitively. Therefore, an invariant sigma-finite
measure $m_n$, which is finite on any bounded set, is defined on the
hypersurface; this measure is the image of the Haar measure on
$SDiag_+(n,\Bbb R)$. In the sequel, it is important that when the
radius is multiplied by a positive number, the invariant measure
also changes being multiplied by the $n$th power of this number,
though it remains an image of the Haar measure. Our aim, as in the
Maxwell-Poincar\'e's Lemma, to find under what conditions the
sequence of the measure spaces $(M^{n-1}_{r_n},\; m_n)$ has a limit
in some sense and to study the properties of the limiting measure.

The difference with the spherical case are rather important. First
of all, in our case the measure $m_n$ is not a probability measure
any more but only a sigma-finite one. Second, the group of
symmetries is commutative while in the spherical case it is the
group $SO(n)$. All this brings us to a different interpretation of
the weak limit. In particular, the manifolds $M^{n-1}_r$ are
embedded into the space of distributions (more exactly -to the cone
of the discrete measures), not into the space of sequences (${\Bbb
R}^{\infty}$) as in the case of spheres.

Notice that the positivity property of the coordinates $x_k$ and of
the group will be lifted in the sequel and we will consider the
whole group $SDiag(n,\Bbb R)$; however, the main point of the
problem will clear up already in this particular case.

\subsection{Embedding of the orbits into the cone of discrete
measures}

The embedding of hyperspheres into the infinite-dimensional vector
space is more complicate than in the case of Maxwell-Poincare lemma
it is not "discrete" but continuous. Let $X$ is the interval $[0,1]$
with the Lebesgue measure $m$, or an arbitrary manifold with the
finite positive continuous measure (or even a measure space, which
is isomorphic to $[0,1]$ with $m$). Let $K(X)$ is the cone of all
finite positive discrete measures: $K(X)=\{\sum_k c_k\delta_{t_k};
c_k>0, \sum c_k < \infty, t_k \in X\}$. The topology on the cone
$K(X)$ will be defined, for instance, as a usual weak topology in
the usual sense duality of the cone $K(X)$ and piecewise constant
measurable functions. It is natural to consider cone $K(X)$ as cone
in the space Schwartz distributions $S(X)\supset K(X)$

 Choose any sequence $\{t_k\}_{k=1}^{\infty}$ which is
\textbf{uniformly (w.r.t. measure $m$) distributed} in $X$, - our
construction depends on the choice of the sequence $\{t_k\}$ but the
final result does not depend. Embed the hypersurface $M^n_{r_n}$
into $K(X)$ sending each vector as follow:
$$y=(y_1, \dots y_n)\mapsto \xi_y=\sum_{k=1}^n y_k \delta_{t_k}.$$
Let $\mu_{n,\theta}$ are the image of the invariant measures on the
manifolds $M^n_r$ under the defined embedding.

\subsection{Weak convergence: Laplace-type definitions. }

We will consider the real Borel finite or sigma-finite measures on
the cone $K$ which take finite values on precompact (= relatively
compact) sets in $K$. Let us introduce a notion of weak convergence
in itself for Borel measures. This can be done in a traditional way
defining the convergence of measures as the convergence of the
integrals on a certain class of functions or sets. Minor
difficulties arise as a result of the infiniteness of measures.
However, we adopt here, for the sake of brevity, a more direct and
convenient way. In what follows, we restrict ourself only with those
measures $\mu$ on the cone $K$ for which the \emph{Laplace
transform} $\hat \mu$ (or the characteristic functional) \emph{is
defined for every step function} $f \in K $:
$${\hat \mu}(f)\equiv
\int_K \exp\bigg\{-\int_X f(x)g(x)dx\bigg\}d\mu(g) <\infty,$$ and,
in accordance with this notion, we assume the following definition.

 \begin{definition}
A sequence of sigma-finite Borel measures $\mu_n$ on the cone $K$ is
said to weakly converge in itself if, for any step function $f \in
K$, the sequence $\lim_n {\hat \mu}_n(f)$ converges; we say that the
sequence $\mu_n$ converges to a measure $\mu$ if the functional
$\lim_n {\hat \mu}_n(f)$ is the Laplace transform of some measure
$\mu$ that is concentrated on the cone $K$ itself, not on its
completion.
 \end{definition}

For finite measures, this definition coincides with the usual one.
Thus we defined a weak limit of the (finite or sigma-finite measures
using Laplace transform.

\subsection{The limit of the Laplace transforms
of the invariant measures on the orbits of Cartan subgroups.}

We will repeat the theorem in a slightly different form and the give
the plan of the proof which based on the direct calculations
\begin{theorem}
Let us choose the radius of our hypersurface $M_{r_{n-1}}^{n-1}$ be
equal to $ r_{n,\theta} \equiv \exp(-\theta n), \theta>0$ and denote
as $\mu_{n,r_{n,\theta}}\equiv \mu_{n,\theta}$ the image of
$SDiag_+(n)$-invariant sigma-finite (uniform) measures $m_n$ on the
hypersurfaces $$M_{r_{n-1}}^{n-1}\equiv M_{n,\theta}= \{(y_1, \dots
y_n): \prod_{k=1}^n y_k =\exp(-\theta n^2); y_k>0, k=1 \dots n \}.$$

 Then the sequence of measures $\mu_{n,\theta}$ on the cone $K$
 weakly converges in itself in the sense of previous definition. In
 another words  the sequence of Laplace transform  ${\hat
 \mu}_{n,\theta}$ of the measures $\mu_{n,\theta}$ converges, and the
 limit is equal to the following functional:

$$\lim_n {\hat \mu}_{n,\theta}(f)=\exp(-\varphi(\theta)\int_X\ln f(x)dx),$$
  where $\varphi(\cdot)$ is a positive function of parameter $\theta >0$.

 The choice of the sequence of the radiuses for which the limit
 exists and does not equal to zero or infinity (as in MP-lemma) is
 unique up to equivalence of asymptotics, and consistent
 normalization of the measures $m_n$ (e.g. up to choice of the set of
 unit measure). Under the embedding of the cone $K$ in the space of
 Schwartz distributions $S(X)$, the sequence of measure
 $\mu{n,\theta}$ converges in a certain sense to a limit measure,
 which we denote by ${\cal L}^+_{\theta}$.
\end{theorem}

There are several plans of the proof of this theorem. The first one
based on the fact to the measure  ${\cal L}^+_{\theta}$ as it was
defined in \cite{GGV,TVY} is invariant under the abelian  group
$\cal M$ of multiplicators (see below) can be applied individual
ergodic theorem (for sigma-finite measures) so individual ergodic
theorem (for sigma-finite measures) can be applied to it, and the
convergence of the finite dimensional approximations is exactly
convergence in ergodic theorem for the integrable functionals
(ergodic method). In this case we already use the existence of the
measures which was proved with different method. Below we will
present of the draft of the direct proofs. We will return to all
this question elsewhere.

 The analogy with Maxwell-Poincar\'e's Lemma consist in the same
 procedure: we calculate the weak limit of the invariant measures on
 the manifolds of the growing dimension under the special choice of
 sequences of "radiuses" of manifolds; but the analogy seems to be
 finished here not because of the big differences between manifolds
 and "radiuses" (in our case the radius is exponentially small and in
 that case is proportional to square root of dimension) but the main
 difference is in the group symmetries - we have noncompact abelian
 group and in the Lemma it was orthogonal group. The limit measure in
 classical case is Gaussian measure and in our case the limit measure
 whose Laplace transform is the right-hand side of the formula above
 needs to be described - it will be done independently on the theorem
 above and we will establish weak convergence under the imbedding
 above to the measure which we will call {\it infinite dimensional
 Lebesgue measure}. We will see that this measure concentrated on the
 Schwartz distributions which is the linear combinations of the
 delta-functions; recall the Gaussian (Wiener) measure is
 concentrated on the Holder functions.

 What does it mean weak convergence? We used ergodic method (or
 weak convergence in the geometrical variant which is not so
 convenient for infinite measures.
 The simplest way to explain the weak convergence (in the theorem above)
 for sigma-finite  measures is to use the convergence of its Laplace transforms.
 The Laplace transform  of measure $\nu$ in the vector space $E$ is
 $$\Psi(f)=\int_E \exp\{-<f,\xi>\} d{\cal L^+_{\theta}}$$

 In our case $\xi=\sum_k y_k\delta_{t_k}$; so we must calculate finite dimensional Laplace
 transform; it is given by integrals with respect to the measure  $\mu_{n,\theta}$
 which is image of the invariant measure $m_n$ on the hyperspheres $M^n_{\theta}=
 \{\{y_k\}: y_k>0, k=1\dots n; \prod_{k=1}^n y_k =\exp(-\theta
 n^2)\}$. Let $f(.)\in K(X)^*$ is
 positive tame function on the manifold $X$ (say, piece-wise constant
 function) Then

 $$D_{n,\theta}(f)=\int (n) \int _{M_{n,\theta}}
 \exp\{-\sum_{k=1}^n y_k \cdot f(t_k)\}dm_n(y).$$

In the following calculations we consider only the case $\theta=1$;
the general case can be easily reduced to it (see below).

Denote $D_n=D_{n,1}$ and $M_n=M_{n,1}$.  Changing the variables
$y_k\mapsto \frac{f(t_k) y_k}{\rho(f)}$,
 where $\rho(f)=(\prod_{k=1}^n f(t_k))^{\frac{1}{n}}\approx \exp{
 \int \log f(t) dm(t)}$, we obtain:
 $$D_n(f)=\int (n) \int _{M_n}
\exp\{-\rho(f)^{-1} \sum_{k=1}^n y_k\} dm_n(y).$$

Let $y_k=e^{x_k},k=1,\dots n$, then our expression equal to
$$=\int (n) \int_{P_n} \exp\{-\rho(f)^{-1}\sum_{k=1}^n
\exp x_k\}\prod_k dx_k,$$ where $P_n=\{(x_1 \dots x_n):\sum_k
x_k=-n^2\}$.

 Finally, change $x_k\mapsto x_k - n$
then we have the following expression for Laplace transform of
measure:
$$D_n(f)=$$
$$=\int_{{\Bbb R}^n} \exp{\{-\rho(f)^{-1} e^{-n}\sum_{k=1}^n
   e^{x_k}\}}\delta_0(\sum_{k=1}^n x_k)\prod_{k=1}^n dx_k\equiv$$
   $$=\int_{H_n}exp\{-\rho(f)^{-1} e^{-n} \sum_{k=1}^n
   e^{x_k}\}dx, $$
where integration is over hyperplane $$H_n=\{(x_1,\dots x_n):\sum_k
x_k=0\}.$$

 Introduce the function:
$$F_n(\lambda)=\int_{H_n}exp \{-\lambda \sum_{k=1}^n \exp x_k\}dx$$
The integration here also takes place over hyperplane
$H_n=\{(x_1,\dots x_n):\sum_k x_k=0\}$ with the Lebesgue measure on.
This is well-known Mellin-Barnes function (Related to Inverse Mellin
transform of Euler Gamma.) It satisfies to the differential equation
\footnote{I grateful to Professor Graev who informed me about this}:
$$(1+\lambda\frac{d}{d\lambda})^{n-1}\frac{dF_n}{d\lambda}=F_n(\lambda)$$

Our calculations gave the following link with Mellin-Barnes
function:

The limit of the Laplace transform depends on the following
characteristic of the argument, tame function $f$: $\rho(f)=\exp
\{\int_T \log f(t)dm(t)\}$ and equal to:
$$D_n(f)=F_n(\rho e^{-n}).$$ where . In other words we need to find \textbf{the
asymptotic of classical Mellin-Barnes function $F_n(\gamma
e^{-\theta n})$ when index $n$ of functions tends to infinity and
argument tends to zero exponentially in $n$.}

The existence of another asymptotics is also interesting question
from the point of view of the measure theory on the infinite
dimensional manifolds. Prof. D.Zagier gave the positive answer on my
 question about the existence of the following limit:

 \begin{statement}
There exist finite limit ,
 $$\lim_{n \to \infty} [F_n(\lambda)]^{1/n}
 \equiv \lim_{n\to \infty} (Mel^{-1}\{\Gamma^n\})^{1/n}\equiv F_{\infty}(\lambda)$$
 for all positive  $\lambda$, where $Mel^{-1}$ is inverse Mellin transform (see \cite{RG})
 \footnote{A detailed calculation of the function $F_{\infty}$
 and further comments on the geometrical meaning of this calculations
 for the infinite-dimensional Lebesgue measure will be published in my paper in forthcoming
 issue (dedicated to V.I.Arnold) of the {\it Journal of Fixed Point Theory and applications,
 vol.3 (2008)}.}
\end{statement}
It seems that the function $F_{\infty}(.)$ had never considered
before. More detail consideration of this subject we postpone till
the next occasion.

\begin{remark}
The characteristic functionals $\Phi_{\theta}$ which we have
obtained are invariant with respect to multiplication of the
arguments on any measurable nonnegative function $a(.)$ with zero
integral of the logarithm:
 $$\Phi(a\cdot f)=\exp\{-\theta\int_X \ln a(x)f(x)dx\}=\exp\{ - \theta [\int_X\ln a(x)dx+\int_X\ln
 f(x)dx]\}= \exp\{-\theta \int_X\ln f(x)dx\},$$
and are invariant up to multiplicative constant if the integral
$\int_X \ln a(x)dx$ is finite.
\end{remark}

Consequently the sigma-finite measure whose Laplace transform is
$\Phi_{\theta}$ must be invariant (correspondingly projectively
invariant) with respect to the group of multiplicators $M_a$ on the
functions $a$   with zero (correspondingly -finite) integral of the
logarithm.

Note that the direct way to establish the weak convergence of the
measures o the orbits consists in the calculations of the
distributions of the finite number of the functionals - this leads
to the weak distribution which we consider in the next paragraph?
nevertheless, to prove a weak convergence for the sigma-finite
measure is not so easy problem as the same fact for the finite
measure, and the notion of the weak distribution for sigma-finite
measures not so natural, this is why  we used Laplace transform.

 Another calculations based on the probabilistic approach.
  Let
$D_{n,\theta}(t\lambda_1,\dots t\lambda_n)\equiv D_{n,\theta}(f)$
where $f$ is a piecewise constant function with values $t\lambda_1,
\dots t\lambda_n$. Then the function $t \mapsto
D_{n,\theta}(t\lambda_1,\dots t\lambda_n)$ for fixed
$\{\lambda_k\}_k$ gives the Laplace transform of the distribution
(with respect to Lebesgue measure) of the sum of exponents

$$\quad \sum_{k=1}^n \lambda_k \exp (x_k-\theta n)$$ under the
conditions: $\sum_{k=1}^n x_k=0$. It is enough to consider the case
$\lambda_k\equiv 1$. In other words we need to find Lebesgue measure
of the set of vectors, which are satisfy to the conditions:

   $$\lim_{n\to \infty} \mbox{Leb}_{n-1}\{(x_1,x_2 \dots x_n):\quad \sum_{k=1}^n
   x_k=0; \quad
   \sum_{k=1}^n e^{x_k} \leq  s e^{\theta n}\}.$$

A comparison with other results suggested to the guess that this
limit for $\theta=1$ must be equal to $Cs$  where $C$ is a constant
which depends on normalization of the Lebesgue measure, but the
author does not know if this is true.

\section{Description of the Lebesgue measures ${\cal L}_{\theta}^+$ and of the Poisson--Dirichlet measure.}

\subsection{Measures ${\cal L}_{\theta}^+$ as weak distributions.}

Now we proceed to the description of the measures which have been described indirectly so far
and which are our main object. We need to prove that, in some completion of the cone $K$,
there exists a one-parameter family of measures ${\cal L}_{\theta}$ with the
following remarkable Laplace transform:
$$\int_K \exp( - \langle f,g\rangle) d{\cal L}_{\theta}^+(g)=\Phi_{\theta}(f)\equiv
\exp\left(- \theta \int_X \ln f(x)dx\right),$$ $\theta>0$,
and to explain what set supports it. For $\theta =1$, this measure ${\cal L}_1^+$ is
the one that should be called \emph{the multiplicative Lebesgue measure in the
infinite-dimensional space}. All these measures are supported by some completion of the cone
$K$, whereas the cone itself has measure zero for all  $\theta$.

First, we describe these measures in a way this is done for weak distributions, namely,
by means of coordinated families of finite-dimensional sigma-finite measures.
For that, we restrict our characteristic functional $\Phi_{\theta}$ to the
finite-dimensional cone of step functions that are constant on the elements of a given
finite partition
$\xi$ of the set $X$, $X=\cup_{k=1}^n F_k$, and take the inverse Laplace transform. As a result
of this direct computation, we obtain some sigma-finite measures $L_{\theta,\xi}$ in ${\Bbb R}^n$
whose densities are described as follows.

\begin{statement}
The density of the measure $L_{\theta,\xi}$ with respect to the Lebesgue measure is
$$\frac{dL_{\theta,\xi}}{dx}(x_1,\dots x_n)=\prod_{k=1}^n \frac{1}
 {\Gamma(\theta m_k)}x_k^{\theta m_k-1},\, x_k>0, \quad k=1, \dots, n$$
(here $m_k$ is the Lebesgue measure of the set $F_k,\; \Gamma(\cdot)$ is the Euler Gamma).
\end{statement}
 See \cite{GGV}, and also \cite{TVY}, where the measures ${\cal L}_{\theta}^+$
were defined in a different way.
\begin{proof}
The formula is checked using the standard formulas for the integrals of gamma distributions.
\end{proof}
We note that the consistency of the measures relative to the refinement of the partitions cannot
be interpreted in the sense of projections of finite-dimensional spaces, as for finite measures:
this is impossible since the projections are infinite. A dual description is involved instead:
the Laplace transforms of all finite-dimensional distributions are the restrictions to
finite-dimensional subspaces of a single functional. The two interpretations of the consistency
are equivalent in the case of probability measures. Specifically, in the case where $\theta=1$,
all these finite-dimensional measures are the Lebesgue measures with consistent normalization
(say, on the unit cubes).

This description is an analog of a pre-measure, or a weak distribution in an
infinite-dimensional vector space, and does not present an explicit description of the
measure itself. However, it helps to see that the corresponding measure (we will see
that it exists) is an analog of the measure generated by the process with independent
nonnegative values, yet a sigma-finite one. We will give a direct description of such measures.

\subsection{Direct description of the measures ${\cal L}_{\theta}^+$
using the Poisson--Dirichlet measures}

Consider another cone
$$D_+=\Big\{\xi=\sum c_i \delta_{x_i},\, x_i \in X,\, c_i>0,\, \sum c_i <\infty\Big\}$$
of all positive finite (non-normalized) measures with countable support in the space $X$.
If $X$ is a segment, such a measure may be regarded as a monotone step function with
countable number of jumps whose sum is finite. In stochastic processes, probability
measures on such a space are called subordinators. We would prefer to regard the elements of
$D_+$ as positive discrete measures, i.e., the positive linear combinations of delta functions,
the more so because the previous interpretation is possible only on a segment.

There is a natural coupling between the space $D_+$ and the cone $K$: each step function
$f=\sum f_k \chi_{F_k}$ defines a functional on $D_+$:
$$
\langle f,\xi\rangle=\sum_k f_k\cdot \Big(\sum_{i: x_i \in F_k}c_i\Big).
$$
Therefore, the cone $D_+$ lies in the weak completion of the cone $K$; we
will not use this later. We define the measures ${\cal L}_{\theta}^+$
on the cone $D_+$ in a direct way and show that they are the continuations of the above-defined
weak distributions on the cone $K$ to true sigma-additive sigma-finite measures.

To do this, we describe the cone $D_+$ in a more convenient and direct way. Namely,
consider the family $\Sigma_{\infty}$ of decreasing (in a nonstrict way) series with
nonnegative summands and finite nonzero sums. This family constitutes a blunted cone
(without the vertex) with an infinite-dimensional simplex $\Sigma_1$ of the monotone
nonnegative series summing to one as a base. Note that
$\Sigma_{\infty}=\Sigma_1\times {\Bbb R}_+$.
Let $X^{\infty}$ be the direct product of a countable number of copies of the space $X$.
We take the product
$$\Sigma_{\infty}\times X^{\infty} =\Sigma_1\times {\Bbb R}_+\times X^{\infty}$$
and identify it with $D_+$ using the map $T$ that sends the pair made up by the series
$\{c_1\geq c_2 \dots \}\in \Sigma_{\infty}$ and the sequence
$\{x_1,x_2, \dots\}\in X^{\infty}$ to a discrete measure as follows:
 $$T\Big(\{c_k\},\{x_k\}\Big)=\sum_k c_k\cdot \delta_{x_k}\in D_+.$$
It is clear that $T$ is a bijection between the product
$$\Sigma_{\infty}\times X^{\infty}$$ and the space $D_+$.

Next we describe the measures on $D_+$ as the $T$-images of some canonical measures.
Take a product measure $m^{\infty}$ (a Bernoulli measure) on $X^{\infty}$
(it does not depend on $\theta$). We consider a one-parameter family of probability
\emph{Poisson--Dirichlet} measures $PD_{\theta},\, \theta>0,$ on the simplex
$\Sigma_1$, see \cite{King}; we discuss them below and in Appendix 1. The most significant
of them, the proper Poisson--Dirichlet measure, corresponds to $\theta=1$. Finally,
we introduce the measures on the half-line $\Bbb R_+$ defined by the density
$dL_{\theta}=\frac{t^{\theta-1}}{\Gamma(\theta)}dt,\, \theta >0,$ relative to the Lebesgue
measure; it is the Lebesgue measure on the half-line if $\theta=1$.

A useful notation for the measure on the cone $\Sigma_{\infty}$ of monotone convergent
positive series is
$$PDC_{\theta}= PD_{\theta}\times L_{\theta}.$$
The measures $PDC$ might be called the ``poissonization'' of the Poisson--Dirichlet measures
(or the conic Poisson--Dirichlet measures), in contrast to the usual measures $PD(\theta)$
concentrated on the simplex $\Sigma_1$. It seems that the sigma-finite measures $PDC_{\theta}$
have not been considered so far.

\begin{definition}
   The measure ${\cal L}_{\theta}^+$ on the cone $D_+$ is defined as the
   $T$-image of the product of measures:

   $${\cal L}_{\theta}^+=T\Big(PDC(\theta)\times m^{\infty}\Big).$$
\end{definition}
It is clear that these measures are sigma-finite, sigma-additive and finite on compact sets.
The following theorem identifies the measure ${\cal L}_{\theta}^+$ and the measure with
Laplace transform equal to the above-computed functional. To be precise, we prove that
this measure corresponds to the weak distribution introduced above and computed in Proposition~2.
Further, this implies that the weak distribution in question leads to the measure with the given
Laplace transform and therefore, by Theorem~2, these measures are the weak limits of the measures
on the orbits.

\begin{theorem}
$$\int_{D_+} \exp\big\{-\langle f,\xi\rangle\big\}\,d{\cal L}_\theta^+(\xi)=
\Phi_{\theta}(f)\equiv\exp\Big\{-\int_X \ln f(x)\,dx\Big\}.$$
Thus the measures ${\cal L}_{\theta}^+$ are the weak limits of the measures on the positive parts of
the Cartan subgroups.
\end{theorem}

\begin{proof}
We use the following remarkable property of the conic Poisson--Dirichlet measures
supported by the cone $\Sigma_{\infty}$.

\begin{theorem}
Consider an arbitrary random partition of the set of positive integers
$\Bbb N$ into a finite number\, $r$ of subsets. In other words, we ascribe each
positive integer, independently of the others, to one of the\, $r$ subsets with
equal probability $1/r$. Then the joint distribution of\, $r$ partial sums
over these sets of a random series (distributed according the
measure $PDC(\theta)$) is the product measure
$\underbrace{L_{\theta}\times\dots\times L_{\theta}}_r$ in ${\Bbb R}^r_+$.
\end{theorem}
We do not prove this characteristic property of the measures $PDC(\theta)$ here.
The corresponding property of the measures $PD(\theta)$, with the multiple product
of measures replaced by the Lebesgue measure on the $r$-dimensional simplex,
follows from the results in \cite{YorP} about the relation between these measures and
the L\'evy processes defined by stable laws; however, it can be deduced directly
from the definitions of these measures (see Appendix). In the sequel, we use only this
characteristic property of the measures $PDC(\theta)$; it shows that the operations
on the measures $PD(\theta)$ are closely connected with the admissible independence
of the the terms of the series. It immediately follows from this property that the measure
${\cal L}_\theta^+$ is a continuation of the weak distribution described in the
previous section, and thus it has the Laplace transform we need.
\end{proof}

The most profound properties of the measures ${\cal L}_\theta^+$, including
their invariance relative to multiplication operators, are evidently related to
the properties of the Poisson--Dirichlet measures. On the contrary, the Poisson--Dirichlet
measures can be defined via the measures ${\cal L}_\theta^+$ as the projections
onto a simplex (ore a cone).

\subsection{Relationship with the gamma process, and a different definition of the measures
${\cal L}_\theta^+$}

Gamma distribution on the half-line $[0,\infty)$ is the distribution
with density $\frac{t^{\theta-1}e^{-t}dt}{\Gamma(\theta)}$ relative to the Lebesgue
measure. This infinitely divisible distribution generates the L\'evy process $y_{\theta}$
with characteristic functional
$$\chi_{\theta}(f)=\exp\Big\{-\theta\int \ln\big(1+f(x)\big)\,dx\Big\}.$$

The realizations of this process, with probability one, are discrete positive measures
with countable support on $X$, i.e., countable linear combinations
$\sum c_k\delta_{x_k}, x_k \in X, c_k>0,\, k=1,2,\dots,$
with finite total charge $\sum_k c_k < \infty$. The distribution of this charge
(i.e., of the sums $\sum_k c_k $) is the gamma distribution. The law of this process will be
denoted by ${\cal G}_{\theta}$.

\begin{theorem}
The measure ${\cal L}_\theta^+$ is absolutely continuous relative to the measure generated by
the gamma process $\chi_{\theta}$, with density
$\frac{d{\cal L}_{\theta}}{d{\cal G}_{\theta}}(\xi)=\exp \big\{\sum_k
c_k\big\}$, where $\xi=\sum_k c_k$. This density is not integrable, due to the
infiniteness of the sigma-finite measure ${\cal L}_{\theta}$.
\end{theorem}

\begin{corollary}
The measure ${\cal G}_{\theta}$ is quasi-invariant relative to the multiplication by functions
with finite integral of the logarithm.
\end{corollary}

Note that in \cite{GGV,TVY}, the statement of this theorem was the definition of the measures
${\cal L}_{\theta}$, thus all properties of ${\cal L}_{\theta}$ were deduced from the properties
of ${\cal G}_{\theta}$. For instance, the invariance relative to the multipliers was deduced
from the quasi-invariance of the measure ${\cal G}_{\theta}$ and the type of the density.
Here we choose an opposite and more natural line (though the proof of the quasi-invariance
of the measure ${\cal G}_{\theta}$ was established in \cite{GGV,TVY} without difficulty):
we use the weak approximation by finite invariant measures and their relation,
important on its own, with the Poisson--Dirichlet measures. Moreover, the remarkable
and characterizing properties of the gamma process find a natural explanation under this
approach.

It was shown in \cite{VY} that the sigma-finite measure ${\cal L}_{\theta}$ may be treated as
a derivative of the infinite-dimensional distribution of the L\'evy processes according to the
parameter $\alpha$ of stable laws at the point $\alpha=0$ (see also \cite{TVY}). At the same time,
to obtain the distribution of the gamma process in a similar way, a passage to the (weak) limit
as $\alpha \to 0$ with simultaneous renormalization of the measures is also needed.
Thus the measure ${\cal L}_{\theta}$ is absolutely continuous relative to the distribution of the
gamma process, but it is more natural to regard it as a derivative with respect to $\alpha$. This fact
is undoubtedly deeply related with the representation theory of the group of
the $SL(2,\Bbb R)$-currents since the state corresponding to the ground representation,
which lies in the base of the construction of the irreducible representation of the current
group (the canonical state), is the exponent of the derivative of the spherical function
corresponding to the complemented series, with respect to the parameter, taken at the end point
(see \cite{GGV}).
This is not a formal resemblance since the above-indicated realization of the representation
is constructed using the measure ${\cal L}_1$ that is a derivative with respect to the same
parameter. The relation of stable laws spherical functions of the complemented series is
doubtless. All this suggests the comparison of the Wiener measure corresponding to $\alpha=2$
with the measure ${\cal L}_1$ corresponding, as was indicated, to $\alpha=0$: these values are
the ends of the segment $[0,2]$ whose points parameterize stable laws. The symmetry group of
these two measures is an infinite-dimensional group of linear transformations in both cases:
the group of orthogonal operators in the Hilbert space in the case of the Wiener measure, and
the commutative group of multipliers in the case of the group of measure preserving
transformations. Stable laws form a sort of deformation joining these two laws; their
symmetry groups (already essentially nonlinear) are not described yet. We may conjecture that
they constitute a nonlinear deformation similar to the homotopy between the orthogonal group
and the diagonal one.

The symmetrized gamma process induced by the symmetric gamma distribution
$\frac{|t|^{\theta-1}e^{-|t|}dt}{2\Gamma(\theta)}$ on the line is similarly related to
the measures ${\cal L}_{\theta}$ introduced below in Subsection 5.1.

\subsection{An additive version of the Lebesgue measure in the infinite-dimensional space}

The measures ${\cal L}_{\theta}^+$ constructed above were invariant under the action
of the multiplication operators. It is more habitual to regard the finite-dimensional Lebesgue
measure as a unique (up to a factor) shift-invariant measure. By taking logarithms of the
elements of the support of the measure constructed, one can transform them into
shift-invariant measures.

Consider the cone $K^+$, see Sect.~3, of positive step functions on $X$ and the measures on it.
We pass from the multiplicative notation of the actions of the multipliers to the additive
one, i.e., we take logarithms of the elements of $K$ and of the multipliers. Then the cone
turns into the vector space $V$ of step functions and the finite-dimensional Cartan groups
$SDiag_+(n,\Bbb R)$ into the vector spaces of dimension $n-1$ that act on $V$ additively.
$V$

We come to the following, probably more transparent, situation. The map $Log$ transforms
the space $D_+$ of discrete positive measures of finite variation into a vector space,
namely, the space $E (X) =\big\{\sum_k b_k\delta_{x_k},\, x_k \in X;\,
\sum_k \exp (-b_k) <\infty \big\} $ of discrete sigma-finite (signed) measures on the segment $X$:
$$
Log: D_X \mapsto E(X)\qquad Log\Big(\sum_k c_k\cdot
\delta_{x_k}\Big)=-\sum_k \log (c_k) \cdot\delta_{x_k};
$$
it is clear that the sequences $b_k$ must grow to infinity fast
enough. This space is the support of the measure ${\bar {\cal
L}}_{\theta} \equiv Log {\cal L}_{\theta}^+$ that is the image of
the measure  ${\cal L}_{\theta}^+$ under the logarithmic map. The
topology on $E(X)$ is also defined as the image of the topology on
$D_+$ under the map $Log$. The measures ${\bar {\cal L}}_{\theta}$
are infinite, sigma-finite and finite on the compact sets in $E(X)$.
Consider the following action of the vector space
$L^1_{\mu,0}(X)=\{f\in L^1_{\mu}(X):\int f=0\}\subset L^1$ on the
space $E(X)$: $$T_f(\sum_k b_k\delta_{x_k})=\sum_k
[b_k+f(x_k)]\delta_{x_k}$$. Both spaces are the spaces of measures:
of absolutely continuous and, correspondingly, countable signed
measures. Therefore, $G(X)$ is also a Banach space of measures. We
restrict ourselves with the measure ${\bar {\cal L}}_(1)$ in $E(X)$,
which will be regarded as a measure in a wider Banach space $G(X)$.

\begin{theorem}
The Banach space $L^1_{\mu,0}(X)$ acts by the operators $T_f, f \in
L^1_{\mu,0}(X)$ on the space $E(X)$ leaving the measure ${\bar {\cal
L}}_(1)$ invariant. More precisely, for any element $f\in
L^1_{\mu,0}(X)$ a set $E_f$ of total ${\bar {\cal L}}_(1)$-measure
exists such that for all $\omega\in E_f, \omega \equiv \sum_k b_k
\delta_{x_k},$ the image $(T_f)(\omega)\equiv \sum_k [b_k+ f(x_k)
\delta_{x_k}]$ lies in $E(X)$ and $T_f$ leaves invariant the measure
${\bar{\cal L}}_(1)$.
\end{theorem}

The theorem follows from the theorem proved in Sect.~4.2
about the invariance of the multiplicative action, i.e., about the conservation of
the measure under the multiplication by a function with zero integral of the logarithm.
The invariance under the shifts by arbitrary elements of the space $L^1_{\mu}(X)$ can be
obtained when one takes the direct product of the measure constructed and the Lebesgue
measure on the line of constants.

\emph{Thus, we have defined a Banach space and a Borel sigma-finite measure on it
which is invariant under the translations by any elements of some infinite-dimensional
closed subspace.}
This is the circumstance that allows us to call this measure an infinite-dimensional additive
Lebesgue measure.

The map $Log$ allows us to analyze the properties of the measure
${\bar {\cal L}}_(1)$ using the properties of the Poisson--Dirichlet
measure $PDC(1)$. The remark in the theorem about the choice of the
set of total measure is essential (see comments in Sect.~2.2
concerning Maxwell-Poincar\'e's Lemma). Recall that $f\in L^1$ is
not an individual function but a class of coinciding $\mod 0$
functions. Therefore, the action by shifts must be understood in the
following sense. Take an individual function $\hat f $ in the class
$f$ which is defined on some set $A_{\hat f} \subset X$ of total
measure and single out those $\omega\in \Phi (X)$ for which $x_k,\,
k =1,2, \dots,$ are in $A_{\hat f}$. Then the formula
$$T_f\Big(\sum_k b_k \cdot \delta_{x_k}\Big)= \sum_k \big(b_k+f(x_k)\big)\cdot \delta_{x_k}$$
determines such an action: the shift of the coefficients in the
configuration $\omega$ by the values of the function $f$ at the
corresponding points. This formula makes sense and is well defined
relative to the change of values $\mod 0$: if $f=f'\mod 0$, then
$T_f=T_{f'}\mod 0$. Nevertheless, there is no set on which all the
shifts would be defined simultaneously. The reason is somewhat
different from that in the Maxwell-Poincar\'e's Lemma example. Here
the action itself for a fixed element $f\in L_1$ is defined as a
class of $\mod 0$ coinciding transformations. It is interesting
that, in addition, the group of shifts is commutative. This is an
algebraic example of an action of a commutative group with invariant
measure which does not admit a simultaneous individualization (of
the point-wise action) of all the elements in the group. See our
comments about invariant measures in Appendix.

\section{Properties and applications of the measures introduced}

\subsection{Removing the positivity condition}

Up to now, we assumed nonnegativity of the parameters of orbits and groups, i.e.,
the positive part $SDiag_+$ of the Cartan subgroup, the positivity of the step functions
forming the cone $K$ and of the multipliers $a(\cdot)$ acting on them, the positivity of the
series forming the simplex $\Sigma_1$ and the cone $\Sigma_{\infty}$, measures on the half-line
($L_{\theta}$), and so on.
The measures ${\cal L}_{\theta}^+$ we constructed were defined on the cone $D_+$
of discrete positive measures on the space $X$.

It is not difficult to lift the positivity restriction and to extend
all the definitions and statements to the real parameter case. We
mention the evident changes. The whole Cartan subgroup $ SDiag
\subset SL(n,\Bbb R)$ replaces its positive part $SDiag_+$; its
entire orbit $M_n=\{(x_1, \dots x_n):\quad |\prod x_k|=r>0 \}$ is
considered (the condition  $x_k>0$ is lifted); the cone $K$ is
replaced by the vector space of all step functions and, finally, we
consider the multipliers with zero or finite integral of the
\emph{modulus} of their logarithm $\int_X \ln|a(x)|\,dx$ instead of
the $\int_X \ln a(x)\,dx$. The measure space is the family of all
absolutely convergent series with decreasing moduli of their
members\footnote{The family of the series where there are members of
equal moduli has zero measure for all the measures considered, thus
the ordering is defined unambiguously on the set of total measure.}
instead of the cone $\Sigma_{\infty}$ of decreasing positive
convergent series, etc. All the proofs and constructions remain
unaltered, the only essential change worth noting concerns the
construction of the measures (Sect.~5.4). As to the definition of
weak distributions, in all places where the measures on the
half-line ${\Bbb R}_+$ or on the cone ${\Bbb R}^n_+$ were
considered, one must extend them to $\Bbb R$ or ${\Bbb R}^n$ using
the multiplication of the cones by $2^n$ vectors $\varepsilon _1
\dots \varepsilon _n$, where $\varepsilon _k=\pm 1$, with the
uniform measure on them. The extension of the measure $PD(\theta)$
from the simplex of positive monotone series summing to one to the
octahedron $O_1$ of all absolutely convergent series with decreasing
moduli summing to one is made in the same way: one takes the direct
product of the Poisson--Dirichlet measure and the uniform (Haar)
measure on the family of infinite sequences of numbers $\pm1$. Then
we take the space $D$ of all discrete measures (charges) of finite
variation on $X$ instead of the cone $D_+$. The isomorphism of the
space $D$ and the product
$$O_1\times {\Bbb R}\times X^{\infty}$$
is constructed using the extension of the map $T$:
$$T\big(\{c_k\},\{x_k\}\big)=\sum_k c_k\cdot \delta_{x_k}\in D_+,$$
with the only difference that $c_k$ may be positive or negative
numbers with finite sum - $\sum |c_k|<\infty$. We denote the
measures obtained on $D$ by ${\cal L}_{\theta},\, \theta >0$
(omitting the subscript $+$). The measure ${\cal L}_1$ corresponding
to $\theta=1$ is called the infinite-dimensional Lebesgue measure.
As above, the following principal result is true.

 \begin{theorem}
The Lebesgue measure ${\cal L}_1$ is the weak limit of measures on complete orbits, and its
characteristic functional has the form
$$\int_{D} \exp\big\{-\langle f,\xi\rangle\big\}d{\cal L}_1(\xi)=\exp\Big\{-\int_X \ln |f(x)dx|\Big\}.$$
The analogous formula can be written in the case of the measures ${\cal L}_{\theta}$:
$$\int_{D} \exp\big\{-\langle f,\xi\rangle \big\}d{\cal L}_\theta(\xi)=\exp\Big\{-\theta\int_X \ln |f(x)dx|\Big\}.$$
 \end{theorem}

The further properties of these measures will be discussed in the next section.
We note that the difference between the positive and the signed versions are not important, and the
theorems about the invariance and uniqueness are proved in the general case in the same way as
in the positive one.

\subsection{Invariance and uniqueness}
\begin{statement}
The above-constructed measures ${\cal L}_{\theta}$ in the vector space $D$
\begin{description}
    \item[1)] are invariant relative to the group $\cal M$ of multipliers $M_a$ by the
    functions $a \in L^0$ with zero integral $\int_X\ln|a(x)|\,dx$; they are also
    projectively invariant, i.e., are multiplied by the constant $\exp\int_X\ln|a(x)|\,dx$
    if this integral is finite;
    \item[2)] are invariant relative to the group ${\frak A}(X)$ of all transformations that
    leave the measure $m$ on $X$ invariant.
\end{description}
\end{statement}

Both propositions follow directly from the definition of these measures. It follows that the
measures ${\cal L}_{\theta}$ are invariant relative the crossed product
${\frak A}(X)\rightthreetimes \cal M$.

It is easy to show (see \cite{TVY}) that the action of the group $\cal M$, and even
of the crossed product, on $(D, {\cal L}_{\theta})$ is ergodic.
\begin{statement}
The list of the measures invariant and ergodic relative to the group
${\frak A}(X)\rightthreetimes \cal M$ is exhausted by the measures ${\cal L}_{\theta}, \theta>0$.
\end{statement}

The measure ${\cal L}_{\theta}$ is concentrated on countable linear combinations of the
delta functions with absolutely convergent series of coefficients. The distribution of the
sum of the coefficients is the Lebesgue measure on the line. The property of the measures
${\cal L}^+_{\theta}$ expressed in Theorem~4 also holds.

Recall that on the space of countable discrete real measures (or on
the space of countable linear combinations of delta-measures), there
exists an ergodic equivalence relation: the equivalence class
consists of the measures with the same support. This equivalence
relation is ergodic in the case of the measure ${\cal L}_{\theta}$.
In other words, the corresponding partition into the classes is
absolutely nonmeasurable. It is, in essence, the partition into the
orbits (mod 0) of the multiplier group action.

It is interesting that the measures whose support consists of discrete measures has so large
infinite-dimensional group of linear symmetries. For comparison, the support of the white noise,
which also has a large symmetry group (see above), consists of distributions rather than
measures.

\subsection{Application to the current group for the group $SL(2,\Bbb R)$}

The main application of the measures constructed is in the current group representations.
This is how they were first discovered in \cite{VGG2}: the $L^2$ spaces with respect to these
measures are the natural Hilbert spaces where the representations of the current groups
can be implemented. Here the invariance of the measure relative to the multiplications
by the elements of the infinite-dimensional diagonal subgroup is used; this fact generalizes
the classical result about the representations of the group $SL(2,\Bbb R)$, namely,
the possibility to extend the representations from the parabolic subgroup to the Cartan
involution, and consequently to the whole current group.

Consider the group of lower triangular matrices with determinant one and elements in the space
of real functions with integrable logarithm of the modulus.
 $$\left(
  \begin{array}{cc}
   a(\cdot)    &     0     \\
   b(\cdot)    & a(\cdot)^{-1} \\
\end{array}
\right)
$$
Note that this group, together wit the involution
$$
\left(
  \begin{array}{cc}
    0 & 1 \\
    1 & 0 \\
  \end{array}
\right)
,$$
generates the whole group $SL(2,\cal F)$, where ${\cal F}=\Big\{f:\int_X \ln
|f(x)|dx <\infty \Big\}$.
\begin{theorem}
Consider the Hilbert space $L^2(D,{\cal L}_{\theta})$ of complex square-integrable
functions on the space $D$ with measure ${\cal L}_{\theta}$.

The unitary operators
$$(U_{a,b}F)(\xi)=\exp\Big\{i\sum c_k b(x_k)+\int_X \ln|a(x)dx|\Big\}F(M^2_a\xi),$$
where $\xi=\sum_k c_k\delta_{x_k}\in D$, $M_a$ is the operator of multiplication
by the function $a$, define an irreducible unitary representation of the above group
of lower triangular matrices that extends to an irreducible representation of the group
$SL(2,\cal F)$. This representation also extends to a unitary representation of the group
${\frak A}(X)$ of transformations of $X$ that leave the measure $m$ invariant.
\end{theorem}

The correctness of the definition and the fact that operators are
unitary of the operators is the consequence of the fact that the
measures ${\cal L}_{\theta}$ are projectively invariant relative to
the group of multipliers, the remaining properties are proved
directly. The formulas that define the involution are given in
\cite{VGG1}, however the principal possibility to extend the
representation to the group $SL(2,\cal F)$ had been proved in
\cite{VGG2} still before the measures ${\cal L}_{\theta}$ were
discovered. Also note that for all $\theta >0$, the representations
are equivalent; therefore, it suffices to consider only the Lebesgue
measure, i.e., the case $\theta=1$. The mentioned commutative model
of the representation of the current group $SL(2,\cal F)$ is a
direct continual analog of the classical representation of the group
$SL(2,\Bbb R)$ in the space $L^2(\Bbb R)$ of functions on the line
(or the projective line) with the Lebesgue measure: the line is
replaced, in a sense, by the continual product of lines, the space
$D$, and the Lebesgue measure on the line by the Lebesgue measure in
the space $D$ introduced here. It is interesting that the space
$L^2(D, \cal L)$ has the structure of metric factorization, i.e., of
a continual tensor product of the $L^2$ spaces, but this metric
factorization is not isomorphic to the Gaussian, i.e., the Fock
factorization, but is isomorphic to the latter as a Hilbert
factorization (see \cite{VTs}).

\subsection{Many dimensional generalization of the Poisson--Dirichlet measures and the
representations of the current groups of the groups $SO(n,1)$}

We considered the measures in the space $D$ of countable real linear combinations of delta
measures on the space $X$ so far. For applications, it is important to broaden the range
of the coefficients and pass to the vector delta measures. We denote by $D^n(X)\equiv D^n$
the vector space of countable linear combinations $\sum_K c_k\delta_{x_k}$ with coefficients
in the Euclidean space ${\Bbb R}^n$ that satisfy the following two conditions:

1) $\sum_k ||c_k||<\infty$;

2) The space $D^n$ is invariant under the action of the point-wise
action of the orthogonal group $SO(n-1)$ and the homothety group in
${\Bbb R}^n$, i.e., it is invariant with respect to the current
group with coefficients in the group $SO(n-1)\times \Bbb R^*$. In
other words, given a linear combination $\sum_k
c_k\cdot\delta_{x_k}\in D^n$, the linear combination $\sum_k
\varepsilon_k \cdot g_k(c_k)\cdot \delta_{x_k}$, with $\varepsilon_k
\in {\Bbb R}^*, g_k \in SO(n-1),\, k=1,2, \dots$, is also in $D^n$.

The topology in $D^n$ is defined in the usual way. Note that the direct product
$\Sigma^n \times X^{\infty} $, where $\Sigma^n $ is the set of the convergent vector series
with members decreasing in the Euclidean norm, is an everywhere dense thick set in the space
$D^n$. A bijection between $\Sigma^n \times X^{\infty}$ and a dense subset of $D^n$ is
constructed as in the case $n=1$: to an arbitrary linear combination $\sum_k
 c_k\delta_{x_k}\in D^n$, where all $||c_k||$ are different, we assign a decreasing in the norm
permutation of the sequence $c_k$ and the corresponding permutation of $x_k$. Let $T$
denote the converse map (defined in the obvious way).

An analog of the measure ${\cal L}_{\theta}$ in the case $n>1$ was defined in \cite{VG,VG1}
by analogy with the case $n=1$. First, we define the vector gamma process with characteristic
functional
$$ \Phi(f)=\exp\Big\{-\theta \int \ln \big(1+||f(x)||^2\big)\,dx\Big\},$$
with subsequent introduction of a density. The geometry of the
measure (asymptotic approach) as well as Poisson--Dirichlet measures
are in no way used under this approach.

Here we define these measures using geometric point of view and
applying again an analog of the Poisson--Dirichlet measures. A
direct analog of the Poisson--Dirichlet measures as measures on the
convergent series hardly exists in the case $n>1$: it is not clear
what does positivity mean, and thus there is no analog of the
simplex of the series. However, there is an analog of the conic
Poisson--Dirichlet measures which we introduce using the
characteristic property of these measures given in Theorem~4. After
that the sigma-finite measures can be defined in the same way as in
the case $n=1$. We restrict ourselves with the case where
$\theta=1$, for brevity.

We define a generalized (conic) Poisson--Dirichlet measure $PDC^n$ in the space $D^n$ as
a measure in the space of convergent vector series with members decreasing in the Euclidean norm
that have the following property: for any partition of the members of the series independently
into an arbitrary finite number $r$ of classes (see Theorem~4) the joint distribution of the
$r$-dimensional vector composed of the sums of these members over the classes is the
$r$-dimensional Lebesgue measure. It follows from the definition that these measures are
spherically (i.e., in the sense of $SO(n-1)^X$) invariant. The uniqueness of such measure
is verified exactly as in the one-dimensional case.
The measure ${\cal L}_1^n$ on the vector space $D^n$ is defined as the $T$-image of the product
$PDC^n\times m^{\infty}$ of measures. The correctness of the definition follows from the fact that
the $PDC^n$-measure of the family of the series that have at least two
members with equal norms is zero.

\begin{theorem}
\begin{enumerate}
    \item The measure ${\cal L}_1^n$ is sigma-finite and take finite values on compact sets.
    \item The Laplace transform of the measure ${\cal L}_1^n$ is the functional
$$\Phi(f)=\exp\Big\{-\int_X \ln||f(x)||\,dx\Big\}.$$
    \item Thus the measure is invariant under the action (by the pointwise multiplication)
    of the elements $a(\cdot)$ of the group of measurable currents with coefficients in
    $SO(n-1) \times \Bbb R^*$ satisfying the condition $$\int \ln ||a(x)||\,dx=0.$$ Moreover,
    it is invariant relative to all changes of the variable $x$ that leave invariant the measure
    $m$.
    \item There is a natural representation in the Hilbert space $L^2(D^n(X), {\cal L}_1^n)$
    of the current group composed by the elements of $O(n,1)^X$
    with finite integral of the modulus of the current.
\end{enumerate}
\end{theorem}

The items 1-3 are proved as in Sections 3-4 for $n=1$. As to the proof of item~4,
see \cite{VG,VG1}.
We only note that the action of the subgroup of the commutative unipotent currents is
realized by the operators of the multiplication of the functionals $h(\cdot)\in L^2$
by the exponent of a linear functional. The action of the subgroup of compact
currents $SO(n-1)^X$ and of the homotheties is described above: it is the action on the
argument of the functional $h(\cdot)$, and this model generalizes the one given in the previous
Subsection~5.3. A similar definition of the Poisson--Dirichlet measures and of the Lebesgue measures
in the infinite-dimensional Hilbert space is also possible. The details will be given in a
forthcoming paper.

\appendix
\section{Appendix}
\subsection{On the Poisson--Dirichlet measures on the space of positive series}

The Poisson--Dirichlet measure $РD(\theta)$ received widespread interest in the 70s on several
reasons (see \cite{King}, \cite{Arr} \cite{VSh}). They are used in combinatorics, partition
theory, population genetics, etc. Here we touch upon the three most spectacular occurrences
of these measures. A deep analysis of the measure $РD(1)$ and of an interesting
Markov chain related to it appeared in the 70s in papers \cite{VSh1, VSh,Vdan}. Although these
papers are mentioned sometimes (however, insufficiently, in our opinion), the deep analysis and the ideas
developed in them, in particular, the reduction to a stationary Markov chain, did not develop
further for the time being.

\textbf{1. The stick breaking process.} Consider a sequence of independent identically
distributed random variables $\xi_1,\,\xi_2, \dots$ on the unit interval with the Lebesgue measure.
We break the interval into parts putting the points
$$x_1=\xi_1,\, x_2= \xi_2(1-\xi_1),\, \dots,\,
x_n=\xi_n\bigg(1-\sum_{k=1}^{n-1} \xi_k\bigg), \dots $$
one by one, so that the interval is finally broken into a
countable number of parts. The corresponding measure on the family of positive series
summing to one is sometimes called the Ewens measure. One gets the Poisson--Dirichlet measure  $PD(1)$
from it by passing to the variational series: each of the initial series is rearranged using the
(random) permutation in the decreasing order of its members. If the Lebesgue distribution
of the variables $\xi_k$ is replaced
by the distribution with density $\frac{1}{\Gamma(\theta)} t^{\theta-1}$
(relative to the Lebesgue measure), then the same procedure leads us to the measure $PD(\theta)$.

\textbf{2. The limiting distribution of the cycle lengths in a random permutation}
(\cite{VSh}, see also \cite{Arr} and references therein). Consider the symmetric group
$S_n$ and assign to each permutation in it the vector of the lengths of its cycles normalized
by the coefficient $n$, in the descending order, i.e., a point in the simplex
$\Sigma_n=\{(x_1 \dots
x_n):\sum_k x_k=1\}$. Denote by $\mu_n$ the image in $\Sigma_n$, under this map, of
the uniform measure on the group $S_n$ and embed the simplices $\Sigma_n$
into the infinite-dimensional simplex
$\Sigma_{\infty}$. The sequence of the measures $\mu_n$ weakly converges to the measure $PD(1)$.
The measures Меры $PD(\theta)$ are obtained using the same procedure if one replaces
the uniform measure on $S_n$ with the measure defined by the density proportional to the
$({\theta-1})$th power of the number of cycles.

\textbf{3. The limiting distribution of the prime divisors of
positive integers} \cite{Bil, Vdan, Arr, Tan}.

Consider the expansion of positive integers into the product of primes arranged in the descending
order,
$$n=p_1\cdot p_2 \dots p_k,\quad p_1\geq \dots \geq p_k>1,$$
and take the vector $\Big(\frac{\ln p_1}{\ln n},
\dots ,\frac{\ln p_k}{\ln n}\Big)\in \Sigma_1$.
If we take the first $N$ positive integers and the uniform distribution on them,
then we obtain a measure on the simplex, and the sequence of such measures
is weakly convergent in $\Sigma_{\infty}$ to the measure $PD(1)$.

Here many questions are left open. Undoubtedly, a mysterious universality of the
measure $PD(1)$ is present in the additive problems of analytical number theory with
infinite number of summands, and in combinatorics. The comprehension of this phenomenon
advanced slowly and did not reach a satisfactory level so far.%
\footnote{The German mathematician K.~Dickman was the first to put, in 1930, the question
on the distribution of the logarithm of the maximal prime divisor. In the 40s, V.L.~Goncharov
(who apparently did not know Dickman's work) studied the distribution of the maximal cycle
length of the random permutation. The understanding of the identity of the two questions came only
in the 80s.}
The technical reason of the universality is that, as mentioned, the summands of a random series
with respect to these measures have, in a sense, the maximal possible independence. A more
accurate meaning of this statement is revealed when one passes from the random series to the
Markov sequence of the quotients of the summands and the remaining sums, see \cite{VSh}.
This explanation is, however, insufficient for the understanding why such independence
occurs in these and many
other examples.%
\footnote{We can add to the discussion initiated by the letter by V.I.~Arnold in \cite{Not}
that the pioneering work \cite{VSh} and paper \cite{Vdan} are tightly related. When the author
was writing paper \cite{Vdan}, he did not know about \cite{Bil}; however,
though short paper it was, \cite{Vdan} contained some statements that were new as compared to
\cite{Bil} and used the results of \cite{VSh}, including the functional equation for the
Dickman--Goncharov density of the distribution. We note that the functional equations for these
densities introduced in \cite{VSh} and \cite{Vdan} are slightly different and are proved in a
different way, but the solutions remain the same, as well as the statement about the invariant
measure for the Markov operator. Thus the quoting of both papers in the reviews about the
Poisson--Dirichlet measures is a necessity.}

The lifting of the measures $PD(\theta)$ (the ``poissonization'') from the simplex to the cone
of positive monotone convergent series $\Sigma_{\infty}$ with the conic Poisson--Dirichlet
measure $PDC(\theta)=PD(\theta) \times L_{\theta}$ plays a no less important role: see its
characteristic property (Theorem~4). This property can be proved directly; moreover, it is
a consequence of the theorem in \cite{YorP} which states that the measures $PD(\theta)$ are
the measures on the set of the trajectory jumps of the gamma process with parameter $\theta$,
i.e., of the L\'evy process constructed by means of the gamma distribution
$\frac{1}{\Gamma(\theta)}t^{\theta-1}e^{-t}dt$ (see Subsect.~4.3).

Some other characteristic properties of these measures are known. One of them was used above,
another is the recently proved in \cite{Zeit} author's conjecture (see an important preliminary
result in \cite{Tsil}): the measure $PD(1)$ is a unique invariant measure on the simplex
$\Sigma_1$ for the Markov chain generated by the merging and subdivision of the summands
of the series. The Poisson--Dirichlet measures find applications also in representation
theory of the infinite-dimensional symmetric group (see \cite{KOV}). All these facts show
a fundamental character of the Poisson--Dirichlet measures. These measures also play a
role in combinatorics and in the problems concerning the series and partitions
which may be compared to that of
Gaussian measures in the theory of vector spaces. The multi-dimensional generalization
of the Poisson--Dirichlet measures was treated in Subsect.~5.4.

\subsection{Restrictions on the groups imposed by the invariance and quasi-invariance of measures}

The fact that a Borel nonzero nonnegative finite or sigma-finite
measure on a separable group that is left-invariant under all shifts
exists only on locally compact groups is the classical theorem by
A.~Weil \cite{We}; it is ``converse'' to Haar's theorem about the
existence of an invariant measure on locally compact groups. Its
most simple and more recent proof uses representation theory. A
slightly stronger result is that the same statement about the
measures is true if they are only quasi-invariant relative to all
(left) shifts. Therefore, in the case of non-locally compact groups,
one can only ask about the (quasi-)invariance of the measure under
the elements of some subgroup of admissible shifts. For any
quasi-invariant measure on a non-locally compact group, this
subgroup must have measure zero; however, this subgroup can be
massive. For probability measures on groups, the subgroup of
admissible shifts (with quasi-invariant measure) may be a Banach or
a Hilbert infinite-dimensional space (for instance, the group of
admissible shifts for the standard Gaussian measure in ${\Bbb
R}^{\infty}$ is $l^2$). Numerous works of probabilistic or
analytical character are devoted to this subject starting with the
1940s. Such measures, according to a rather improper tradition, are
called quasi-invariant; nevertheless, this does not raise a
confusion because there exist no ``true'' quasi-invariant measures
(i.e., the measures for which the set of admissible shifts has
positive measure). Of special interest are the quasi-invariant
measures on non-Abelian infinite-dimensional groups, which remain
still not adequately studied. They are needed for the development of
the analysis and the representation theory of such groups, and their
applications to theoretical physics (a groups of diffeomorphisms,
current groups, automorphism groups of various structures).

If one wishes that nonnegative and nonzero measure were invariant, rather than quasi-invariant,
with respect to the shifts by the elements of a non-locally compact group, then this measure
must already be infinite. Only sigma-finite Borel measures that take finite values on
compact sets are of interest for us. It is easy to present such examples with meagre group of admissible
shifts. Here is one of them. Consider the infinite product $m^{\infty}$
of the infinite number of copies of the Lebesgue measure $m$ on the unit interval
in the space of all real sequences ${\Bbb R}^{\infty}$,
and a sigma-finite measure that is obtained using
the shifts of this product measure by the
finite integer-valued sequences. This measure is invariant under the translations by finite
vectors in the space ${\Bbb R}_{\infty}$. However, this example is not very interesting
due to the poor family of linear symmetries of the measure. The group of admissible shifts
is merely the sum of finite-dimensional spaces here.

Our example in Subsect.~4.4 of an additive infinite-dimensional Lebesgue measure $Log {\cal L}^+$
is new and unexpected in this very  aspect: \emph{the group of admissible shifts that leave
invariant some sigma-finite measure that is finite on compact sets is an infinite-dimensional
Banach space} ($L^1(X)$). Moreover, this measure is concentrated on the set of countable
linear combinations of delta functions. Possibly that in essence this example exhausts all
the possibilities where the group of shifts is a Banach space. It is interesting, which
non-Abelian complete infinite-dimensional groups can play the role of the group of admissible
shifts. One may expect that the study of such examples would lead to interesting applications
in the theory of infinite-dimensional integration.

\subsection{The model of continuous tensor product which is associated with infinite
dimensional Lebesgue measure} The measure  ${\cal L}_1^n$ for all
values of $n$ gives new model of the continuous tensor product of
the Hilbert space. Usually the right meaning of continuous tensor
product plays Fock space (or exponent of Hilbert space). It is
possible to substitute Fock space with another space $L^2$ over the
law of Levi processes. Using the measure  ${\cal L}_1$ we can give
decomposition of the continuous tensor product onto direct integral
with respect to ${\cal L}_1$ of the countable tensor product of
Hilbert space. More precisely, it is possible to give exact
interpretation of the left side of the formula (continuous tensor
product)
$$
\int_X^{\bigotimes} L^2({\Bbb R}; K)
dm=\int_{D(X)}^{\bigoplus}\bigotimes_{i=1}^{\infty}H_{\xi_i}d{\cal
L}_1(\xi),
$$
using right side of this formula; - here $X$ is an arbitrary
Lebesgue space with finite measure $m$; the space $L^2(\Bbb R; K)$
is a space of $K$-valued $L^2$-functions with respect to Lebesgue
measure on $\Bbb R$ with some auxiliary Hilbert space $K$;
$H_{\lambda}, \lambda \in {\Bbb R}_+$ is a family of Hilbert spaces
which depend of real positive parameter $\lambda$, and related to
the space $K$, and $\xi=\{\xi_i\}$ runs over the elements of the set
of full ${\cal L}_1$-measure in the space $D(X)$. Thus this formula
reduces (or gives definition) of the continuous tensor product (LHS)
to the direct integral of countable tensor products (RHS). The role
of measure ${\cal L}_1$ here is crucial, - we use the invariance and
ergodicity of the measure ${\cal L}_1$ with respect to the group of
multiplicators (see 5.2). One concrete example of such
interpretation will be done in the paper \cite{VG2} concerning to
the representations of the current groups with coefficients in the
groups $O(n,1)$ and $U(n,1)$.

\end{document}